\documentstyle[epsf,11pt,psfig]{article}
\begin{document}
\thispagestyle{empty}
\hspace*{10.0cm}                        WU-B 95-31\\
\hspace*{10.0 cm}                     October 1995\\
\begin{center}
{\Large\bf SPIN EFFECTS IN HARD EXCLUSIVE REACTIONS} \\
\vspace*{1.0 cm}
\end{center}
\begin{center}
{\large P. Kroll}
\footnote{Invited talk presented at
           the SPIN95 Conference, Protvino (1995)}
                 \\
\vspace*{0.5 cm}
Fachbereich Physik, Universit\"{a}t Wuppertal, \\
D-42097 Wuppertal, Germany\\[0.3 cm]
\end{center}
\vspace*{4.0 cm}
\begin{center}
                    {\bf Abstract}
\end{center}
The present status of applications of perturbative QCD to large
momentum transfer exclusive reactions is discussed. It is argued that
in the region of momentum transfer accessible to present day
experiments soft contributions dominate the exclusive processes.\\

\newpage
\setcounter{footnote}{0}
\setcounter{page}{1}
\begin{center}
{\Large\bf SPIN EFFECTS IN HARD EXCLUSIVE REACTIONS} \\
\end{center}
\begin{center}
{\large P. Kroll}
\footnote{E-mail: kroll@wpts0.physik.uni-wuppertal.de}\\
\vspace*{0.1 cm}
Fachbereich Physik, Universit\"{a}t Wuppertal, \\
D-42097 Wuppertal, Germany\\[0.3 cm]
\end{center}
\vspace*{-0.5 cm}
\begin{center}
                    {\bf Abstract}
\end{center}
\vspace*{-0.2 cm}
The present status of applications of perturbative QCD to large
momentum transfer exclusive reactions is discussed. It is argued that
in the region of momentum transfer accessible to present day
experiments soft contributions dominate the exclusive processes.
\vspace*{-0.4cm}
\section{Introduction}
\setcounter{equation}{0}
\vspace*{-0.2cm}
\indent
This talk is not meant as a review but rather represents my opinion about
the current status of the theoretical understanding of hard exclusive
reactions. Although we are mainly interested in spin physics at this
conference I am compelled to discuss unpolarized observables first
since their understanding is prerequisite for a reliable treatment
of the in general smaller spin effects.
\par
\indent
There is general agreement that perturbative QCD in the framework of the
hard-scattering approach (HSA) \cite{lep:80}
is the correct description of form factors and perhaps other exclusive
reactions
at asymptotically large momentum transfer. In that approach a form factor
or a scattering amplitude is expressed by a convolution of distribution
amplitudes (DA) with hard scattering amplitudes to be calculated in collinear
approximation within perturbative QCD. The universal, process independent
DAs, which represent hadronic wave functions integrated over
transverse momenta ($k_{\perp}$), are controlled by long distance physics
in contrast to the hard scattering amplitudes which are governed by
short distance physics. Explicitly,  a helicity amplitude for a
process $AB\to CD$ reads
\begin{eqnarray}
\label{helamp}
{\large M}_{CDAB}(s,t)\,=\,\int \prod_{i=A,B,C,D} [{\rm d}x_i] & &
\phi^*_C(x_C,\mu_F) \phi^*_D(x_D,\mu_F) T_H(x_i,s,t,\mu) \nonumber \\
& & \times \phi_A(x_A,\mu_F) \phi_B(x_B,\mu_F)
\end{eqnarray}
where helicity labels are omitted for convenience. $[{\rm d}x_i]$ is
short for
\nopagebreak[4]
\begin{equation}
\label{ind}
[{\rm d}x_i]\,=\, \prod ^{n_i}_{j=1} {\rm d}x_{ij} \delta(1-\sum x_{ij})
\end{equation}
\nopagebreak[4]
$x_{ij}$ is the momentum fraction the constituent $j$ of hadron $i$
carries.
\nopagebreak[4]
$\mu_F$ is the
\pagebreak
scale at which short and long distance physics
factorizes and $\mu$ is the renormalization scale. If one of the external
particles is point-like, e.~g.~a photon, the corresponding DA is to be
replaced by $\delta (1-x_{i1})$.
\par
\indent
The HSA possesses two characteristic properties, the quark counting rules
and the helicity sum rule. The first property says that the fixed
angle cross-section behaves at large Mandelstam $s$ as
\vspace*{-0.3cm}
\begin{equation}
\label{a1}
\vspace*{-0.2cm}
{\rm d}\sigma /{\rm d}t = f (\theta)\, s^{2-n} \quad\quad ({\rm modulo}\,\,
\log{s})
\vspace*{-0.1cm}
\end{equation}
where n is the minimum number of external particles in the hard
scattering amplitude. The counting rules also apply to form factors:
a baryon form factor behaves as $1/Q^4$, a meson form factor as $1/Q^2$
at large momentum transfer $Q$. These counting rules are in surprisingly good
agreement
with experimental data. Even at momentum transfers as low as 2 GeV
the data seem to respect the counting rules.
\par
\indent
The second characteristic property of the HSA is the conservation of
hadronic helicity. For a two-body process the helicity sum rule reads
\vspace*{-0.3cm}
\begin{equation}
\label{a2}
\vspace*{-0.2cm}
\lambda_A + \lambda_B = \lambda_C + \lambda_D.
\vspace*{-0.1cm}
\end{equation}
It appears as a consequence of utilizing the
collinear approximation and of dealing with (almost) massless quarks
which conserve their helicities when interacting with gluons.
The collinear approximation implies
that the relative orbital angular momentum between the constituents
has a zero component in the direction of the parent hadron. Hence the
helicities of the constituents sum up to the helicity of their parent hadron.
As will be discussed below hadronic helicity is not conserved;
the ratio of hadronic helicity flip matrix elements to non-flip ones is
about 0.2.
\par
\indent
Many hard exclusive reactions have been analysed
within the framework of the HSA: Electromagnetic form factors
of mesons and baryons, Compton scattering off nucleons,
photoproduction of mesons, two-photon annihilations into pairs of mesons
or baryons, decays of heavy mesons etc. No clear picture
has emerged as yet; there are successes and failures. It however
seems that results of the order of the experimental
values are only obtained if, at least for the proton and the pion, DAs
are used which are strongly concentrated in the end-point regions.
Chernyak and Zhitnitsky (CZ) \cite{CZ:82} claimed that such DAs find a certain
justification in QCD sum rules by means of which a few moments of
the DAs have been calculated. The CZ moments are subject of considerable
controversy: Other QCD sum rule studies as well as lattice gauge
theory provide other values for the moments. The asymptotic forms of
the DAs ($\sim x_1 x_2$ for the pion, $\sim x_1 x_2 x_3$ for the
proton), into which any DA evolves for $Q\to\infty$, lead to results
which are typically orders of magnitudes too small as compared with data.
Consider as an example the magnetic form factor of the proton. For the
DAs of the CZ type one obtains $Q^4 G_M \simeq 1 \rm{GeV}^4$ in agreement
with experiment, whereas a vanishing result
is found for the asymptotic DA.
\par
\indent
Purely hadronic reactions, as for instance elastic proton-proton
scattering,
have not yet been studied in the frame work of the HSA.
The reason for that fact is, on the one hand, the huge number of
Feynman graphs contributing to such reactions and, on the other hand,
the occurence of multiple scatterings (pinch singularities \cite{lan:74}),
i.e. the possibility that pairs of constituents scatter independently
in contrast to the HSA in which all constituents collide within a small
region of space-time. A general framework for treating multiple
scattering contributions has been developed by Botts and Sterman \cite{bot:89}.
\vspace*{-0.3cm}
\section{The modified perturbative approach}
\setcounter{equation}{0}
\vspace*{-0.2cm}
\indent
The applicability of the HSA at experimentally accessible momentum
transfers, typically a few GeV, was questioned \cite{Isg:89}. It
was asserted that in the few GeV region the HSA accumulates large
contributions from the soft end-point regions, rendering
the perturbation calculation inconsistent. This is in particular the case
for the end-point concentrated DAs. Another theoretical defect
is caused by the collinear approximation: The neglect of the
transverse momentum dependence of the hard scattering amplitude
is a poor approximation in the end-point regions. The magnitude of the
errors in the final results for, say, the pion's or the nucleon's form
factor, generated by the collinear approximation depends on the shape
of the DAs. Obviously, the CZ-like DAs entail large errors in contrast
to the asymptotic DA and similar forms for which the errors are
sufficiently small.
\par\indent
Recently a modification of the HSA was suggested \cite{bot:89,LiS:92}
in which the transverse momentum dependence of the hard scattering
amplitude is retained and Sudakov corrections are taken into account.
Let me discuss the characteristics of that approach on the example
of the $\pi\gamma$ transition form factor which is written as \cite{jak:94}
\begin{equation}
F_{\pi\gamma}(Q^2)=
\int {\rm d}x \frac{{\rm d}^{\;\!2}b}{4\pi} \; \hat \Psi_0(x,-{\bf b},\mu_F) \;
\hat T_H(x,{\bf b},Q) \; \exp\left[ -S(x,b,Q) \right],
\label{eq:pigaff-ft}
\end{equation}
where ${\bf b}$ is the quark-antiquark separation in the transverse
configuration space. $\hat T_H$ denotes the Fourier transform of the
momentum space hard scattering amplitude.
Neglecting masses, it reads at lowest order (QED)
\begin{equation}
\hat T_H(x,{\bf b},Q) =
\frac{2}{\sqrt{3}\,\pi}
K_0(\sqrt{(1-x)Qb})\,+\,\cal O (\alpha_S) ,
\label{eq:pigaff-TH-ft}
\end{equation}
where $K_0$ is the modified Bessel function of order zero.
The Sudakov exponent $S$ in (\ref{eq:pigaff-ft}) comprising those
gluonic radiative corrections not taken into account by the evolution
of the wave function, is given by
\begin{equation}
S(x,b,Q)= s(x,b,Q)+s(1-x,b,Q)
-\frac{4}{\beta_0} \ln \frac{\ln (\mu/\Lambda_{QCD})}
{\ln (1/b \Lambda_{QCD})}
\label{eq:pigaff-sudakov}
\end{equation}
where a Sudakov function $s$ appears for each quark
line entering the hard scattering amplitude. The last term in
(\ref{eq:pigaff-sudakov}) arises from the application of the renormalization
group equation ($\beta_0=11-2/3 \,n_f$). A value of $200\,{\rm MeV}$ for
$\Lambda_{QCD}$ is used throughout and the renormalization scale $\mu$
is taken to be the largest mass scale appearing in $\hat T_H$,
i.~e., $\mu=\max(\sqrt{1-x}Q,1/b)$. For small $b$ there is no
suppression from the Sudakov factor; as $b$ increases the Sudakov
factor decreases, reaching zero at $b=1/\Lambda_{QCD}$. For even
larger $b$ the Sudakov factor is set to zero. The Sudakov function $s$
is explicitly given in \cite{bot:89,LiS:92}. $b$ plays the r$\hat o$le
of an infrared cut-off; it sets up the interface between
non-perturbative soft gluon contributions -- still contained in the
hadronic wave function $\hat {\Psi}_0$ -- and perturbative soft
gluon contributions accounted for by the Sudakov factor. Hence, the
factorization scale $\mu_F$ is taken to be $1/b$.
\par\indent
The quantity $\hat \Psi_0$ appearing in (\ref{eq:pigaff-ft}) represents
the soft part of the transverse configuration space pion wave function, i.~e.,
the full wave function with the perturbative tail removed from
it. The wave function is parameterized as \cite{jak:93}
\begin{equation}
\hat\Psi_0 (x,{\bf b},\mu_F) = \frac{f_\pi}{2 \sqrt{6}}
\,\phi(x,\mu_F) \,\hat\Sigma(\sqrt{x(1-x)}\,b).
\label{eq:wvfct-ansatz}
\end{equation}
It is subject to the auxiliary conditions
\begin{equation}
\hat\Sigma ( 0 ) =4\pi, \qquad\qquad
\int_0^1 {\rm d}x \;\phi(x,\mu_F) =1.
\label{eq:auxiliary}
\end{equation}
$f_\pi\,(=130.7\,{\rm MeV})$ is the usual $\pi$-decay constant.
The wave function does not factorize in $x$ and $b$, but
in accord with the basic properties of the HSA \cite{zhi:93,lep:83}
the $b$-dependence rather appears in the combination
$\sqrt{x(1-x)}\,b$. The transverse part of the wave function
is assumed to be a simple Gaussian (see \cite{zhi:93} for a discussion
of this ansatz)
\begin{equation}
\hat\Sigma = 4\pi \;
\exp \left[ -x(1-x)\,b^2/4 a^2 \right].
\label{eq:Sigma-ft}
\end{equation}
The transverse size parameter $a$ is fixed for a given DA from the
$\pi^0\to\gamma\gamma$ decay providing the relation \cite{lep:83}
\begin{equation}
\int{\rm d}x\,{\rm d}^{\;\!2}b \; \hat\Psi_0(x,{\bf b},\mu_0)=\sqrt{6}/f_\pi.
\label{eq:pimunupigamgam}
\end{equation}
There is no other parameter to adjust. Utilizing two frequently
used DAs , namely the asymptotic one
\begin{equation}
\label{eq:DA-AS}
\phi_{AS}(x)=6\, x(1-x)
\end{equation}
and a form proposed by Chernyak and Zhitnitsky \cite{CZ:82}
\begin{equation}
\label{eq:DA-CZ}
\phi_{CZ}(x)=30\, x(1-x) \, (1-2x)^2,
\end{equation}
Jakob et al.\cite{jak:94} evaluated the $\pi\gamma$ form factor. The results
are compared to CELLO \cite{Beh:91} and CLEO \cite{sav:95} data in Fig.~1.
The predictions obtained with the asymptotic wave function are in perfect
agreement with the data whereas the CZ wave function, i.~e.~the CZ DA
multiplied by the Gaussian (\ref{eq:Sigma-ft}), leads to results in dramatic
conflict
with the data. In the perturbative approach the
wave function is an universal, i.~e.~process-independent object. Hence, in
analyses of other hard exclusive reactions the AS wave function (or
eventually slightly modified versions of it) should be applied. The
use of the CZ wave function, on the other hand, appears to be
inconsistent in the light of the observations made in \cite{jak:94}.
\par\indent
Applications of the modified perturbative approach to the $\pi$'s and the
nucleon's
electromagnetic form factors \cite{jak:93,bol:95a} reveals that the
perturbative contributions, although self-consistently
evaluated, are too small. In Fig.~2 the results for
the magnetic form factor of the proton are shown \cite{bol:95a}. The
CZ-like DAs \cite{ber:93} lead at best, namely when the wave
functions are normalized to unity, to perturbative results amounting
to about $30\%$ of the experimental values for $Q^2\geq 10 {\rm GeV}^2$.
\vspace*{-0.3cm}
\section{Soft contributions}
\setcounter{equation}{0}
\vspace*{-0.2cm}
\indent
The above mentioned results on form factors clearly indicate the
dominance of soft physics, i.~e.~of higher twist contributions, in the
experimentally accessible region of momentum transfer. At this point
I have to recall the following fact: The elastic form factors also
receive contributions from the overlap of the initial and final state
wave functions $\hat \Psi_0$ \footnote{ Note that formally the perturbative
contribution to elastic form factors represents the overlap of the large
transverse
momentum tails of the wave functions.}.
In the case of the pion's electromagnetic form factor, the overlap
contribution reads in the transverse configuration space
\begin{equation}
{F_\pi}^{soft}(Q^2)=
\frac{1}{4\pi}\int {\rm d}x {\rm d}^{\;\!2}b
\exp\left[i(1-x){\bf b}\cdot {\bf q}_\perp\right]\;
|\hat\Psi_0(x,{\bf b})|^2,
\label{eq:piff-soft-ft}
\end{equation}
where $Q^2={\rm q}_\perp^{\;\!2}$. As shown in \cite{jak:94,jak:93}
the AS wave function provides an overlap contribution of the right
magnitude to fill in the gap between the perturbative contribution
and the and the experimental data. As required by the consistency
of the entire picture, $F_\pi^{soft}$ decreases faster with
increasing $Q$ than the perturbative contribution.
The broad flat maximum of the overlap contribution in the few GeV
region simulates the $Q$-dependence predicted by the dimensional counting
rules. For the CZ wave function the overlap contribution exceeds the
data significantly; the maximum value of $Q^2 F_\pi$ amounts to about
$2.1 \,{\rm GeV}^2$. This result is to be considered as a serious failure of
the CZ wave function. Soft contributions of the overlap type have also
been discussed in \cite{Isg:89,kis:93,dor:95} and observations have
been made similar to those in \cite{jak:94,jak:93}. In the case of
the $\pi\gamma$ transition form factor the overlap contribution
is expected to be very small due to a helicity mismatch.
\par\indent
In Ref.~\cite{bol:95b} the overlap contribution to the nucleon's
magnetic form factor is evaluated where again a Gaussian $b$ dependence
analogue to (\ref{eq:Sigma-ft}) is employed. A rather small overlap
contribution is obtained with the asymptotic wave function
($\phi_{AS}=120x_1x_2x_3$) for the proton and a strict zero for the
neutron. The CZ-like wave functions given in \cite{ber:93}, on the
other hand, lead to very large overlap contributions exceeding the
data by huge factors. Again, as in the case of the pion, the strongly end-point
concentrated wave functions while providing large but theoretically
inconsistent leading twist contributions, have to be rejected. In
order to find a proper wave function for the nucleon the following
strategy is adopted: Starting point is the expansion of the nucleon's DA
over the eigenfunctions of the evolution equation which are
appropriate linear combinations of the Appell polynomials truncated at
the same order as the CZ-like DAs \cite{ber:93}
\begin{equation}
\label{evo}
\phi(x,\mu_F)=120 x_1x_2x_3 [1+\sum_{n=1}^5 B_n(\mu_F) \tilde {\phi}_n(x)].
\end{equation}
The expansion coefficients $B_n$ and $f_N$, playing the
r$\hat{o}$le of the wave function at the origin of the configuration
space, as well as the transverse size parameter are fitted to the
experimental data of the proton's magnetic form factor, the decay
width for $\Psi\to p\bar{p}$ and the inclusive valence quark distribution
functions. Thereby the form factor is evaluated from the overlap
contribution whereas the $\Psi$ decay is calculated within the
modified perturbative approach. For good reasons the dominance of the
perturbative contribution is to be expected for the $\Psi$
decay. Indeed fair agreement with the experimental value for the decay
width is found.
The fitted parameters of that new nucleon wave
function are compiled in Table 1 where, for comparison, also the
\begin{table}[b]
Table 1: $f_N$ and the expansion coefficients $B_i$, $i=1-5$ for the
AS, the COZ\cite{coz:89} and the fitted wave functions\cite{bol:95b} at a
factorization scale $\mu_F$ of $1\,{\rm GeV}$.
\newcommand{\hoe}[1]{\rule{0cm}{#1mm}}
 \begin{center}
   \begin{tabular}{|l||l|r|r|r|r|r|} \hline
      DA \hoe{4}      & $f_N$ [GeV$^2$] & $B_1\;\;$ & $B_2\;\;$ & $B_3\;\;$
                            & $B_4\;\;$ & $B_5\;\;$
      \\ \hline\hline
      asympt. \hoe{4} & $5.0000 \cdot 10^{-3}$  &  0.0000 & 0.0000 &  0.0000 &
                 0.0000 &  0.0000 \\ \hline
      COZ     \hoe{4} & $5.0000 \cdot 10^{-3}$  &  3.6750 & 1.4840 &  2.8980 &
                -6.6150 &  1.0260 \\ \hline
      Fit     \hoe{4} & $7.0671 \cdot 10^{-3}$  & -0.1081 & 0.5167 & -0.2879 &
                -3.8884 & -0.0371 \\ \hline
   \end{tabular}
 \end{center}
\end{table}
parameters for the asymptotic and the COZ \cite{coz:89} wave functions are
shown.
The transverse size parameter has a value of $0.75\,{\rm GeV}^{-1}$
for the fitted wave function. In Fig.~3 the overlap contributions are
compared to the data for the proton and the neutron
magnetic form factors. Obviously the magnitudes of the overlap
contributions are correct in the $Q^2$ range from about $8$ to
$20\,{\rm GeV}^2$.
\nopagebreak[4]
For smaller values of $Q^2$ similar contributions
from higher Fock states are expected to be important
(for the fitted wave function the probability of the
valence Fock state amounts to 0.205).
For $Q^2$ larger than about $20\,{\rm GeV}^2$ the fit is slightly below the
data. Little modifications of the $b$-dependence (\ref{eq:Sigma-ft}) may
cure that defect \cite{zhi:93}.
\vspace*{-0.3cm}
\section{Spin effects}
\setcounter{equation}{0}
\vspace*{-0.2cm}
\indent
As mentioned in the introduction a characteristic property of the HSA
is the conservation of hadronic helicity. Yet the helicity sum rule
(\ref{a2}) is violated at moderately large momentum transfer. For
instance, the single spin asymmetry in proton-proton elastic
scattering amounts to about $20\%$ at a momentum tranfer of
$6\,{\rm GeV}^2$ \cite{cam:85}. Another example is the Pauli form factor of the
proton
which is measured to be very large \cite{bos:92}. Its $Q^2$ dependence
is seen to be compatible with a higher twist contribution
($\sim1/Q^6$), see Fig.~4.
Charmonium decays into proton and antiproton also show evidence for
violations of the helicity sum rule. Consider the spin zero
states, the $\eta_c$ and the $\chi_0$. In the rest frame of the mesons
the decay products, proton and antiproton, must have the same
helicity in obvious contradiction to the helicity sum rule. Hence,
at leading twist, the widths for these two decays are zero. For the
other three reactions, $\Psi, \chi_1, \chi_2 \to p \bar p$, the
leading twist analysis provides non-zero decay widths which, at least for the
COZ DA
\cite{coz:89} are in apparent agreement with the data leaving aside
the theoretical difficulties with the strong end-point contributions
\cite{CZ:82}
\footnote{The $\alpha_S$ values used in these analyses are
  typically too small as compared with the expectations for a
  characteristic mass scale of the order of the charm quark mass
  and a typical value of $200\,{\rm MeV}$ for $\Lambda_{QCD}$. Since the
  decay widths are proportional to $\alpha_S^6$ a large factor of
  uncertainty is therefore hidden in these calculations. In the modified
  perturbative analysis \cite{bol:95b} of the $\Psi$
  decay the average value of $\alpha_S$ has the reasonable value of 0.41.}.
The leading twist pattern does, however, not match with the experimentally
observed pattern of branching ratios
($BR(\eta_C\to p\bar{p})=(1.2\pm 0.4)\cdot 10^{-3}$;\,
$BR(\chi_0\to p\bar{p})\le 0.9\cdot 10^{-3}$ \cite{pdg:94}).
I would like to mention that the E605 collaboration is going to measure,
hopefully precisely enough, the decay $\chi_0\to p \bar{p}$ at FERMILAB.
It will be very interesting to see whether the measured width is close
to the present day upper bound \cite{pdg:94} which would indicate
a strong spin effect, or much smaller as helicity conservation demands.
\par \indent
How can we understand these large violations of hadronic helicity
conservation in perturbation theory?
The simplest idea to generate helicity flips is to take into
account quark masses.
According to Efremov and Teryaev \cite{efr:82}
the relevant mass should be of the order of the hadron mass since
helicity flips are of twist three type. Model builders take this result
as justification for the use of constituent quark masses. Still this
mechanism does not lead to sizeable spin effects. For example, only
values of the order of $10^{-6}$ are obtained for the
$\eta_C,\chi_0\to p \bar{p}$ branching ratios \cite{ans:92}.
\par\indent
Another and perhaps more appealing idea is to take into account
non-zero orbital angular momentum in the wave functions. In this case the
hadron spin fails to equal the sum of the quark spins; a basic
presumption in the derivation of the helicity sum rule is not satisfied. For
example
the pion wave function (including its spin part) may be written as
\begin{equation}
  \label{orb}
\hat \Psi _0(x,{\bf b})\, p\hspace*{-0.2cm}/\,\,\gamma_5\,+\,
\hat \Psi_1(x,{\bf b}) [p\hspace*{-0.2cm}/\,, {\bf
b}\hspace*{-0.2cm}/\,]\,\,\gamma_5.
\end{equation}
${\bf b}$ represents one unit of orbital angular momentum
\footnote{Dorokhov \cite{dor:95} has constructed such a pion wave
  function from the helicity and flavour changing instanton force.}.
In combination with a ${\bf b}$ dependent hard scattering
amplitude such a wave function can provide violations of helicity conservation
within the modified perturbative approach.
Recently, Gousset, Pire and Ralston \cite{gou:95} applied this idea to
elastic scattering. They argue that the multiple scattering mechanism
\cite{lan:74,bot:89} provides the ${\bf b}$ dependent hard scattering
amplitude since the transverse distance between the various elementary
scattering planes provide a characteristic direction. Hence, helicity
non-conservation is obtained in the large $Q^2$ limit without flipping
a quark helicity. However, the multiple scattering mechanism only provides
violations of the helicity sum rule by two units (``two flip rule''); single
spin
asymmetries like that one in $pp$ elastic scattering, remain
unexplained. So far Gousset et al. applied their mechanism only to $\pi\pi\to
\pi\pi$ and $\pi\pi\to\rho\rho$ . As a consequence of the two
flip rule the amplitude for $\pi\pi\to\rho_0\rho_+$ is zero while that
of $\pi\pi\to\rho_+\rho_+$ is non-zero. There
is an objection against the mechanism proposed by Gousset et al: The
multiple scattering contribution only amounts to a small fraction
of the $pp\to pp$ cross section
\cite{bot:91,jak:94b}. Hence, any particular property of it is perhaps not
relevant to what we may see in data.
\par\indent
Last I want to discuss briefly the contribution of the Wuppertal group
to that field. For baryons one may think of quark-quark correlations in the
wave
functions which also constitute higher twist effects. In a series of
papers \cite{jkss:93,kro:93,pil:93,gui:95} we have advocated that correlations
of this type can effectively be described as quasi-elementary
diquarks. That diquark model is a variant of the HSA: The valence
Fock state of baryons consists of quark and diquark
with a corresponding DA and the hard scattering amplitude is
the amplitude for a subprocess involving quarks and diquarks. In order
to take care of the composite nature of the diquarks and in order to
guarantee that asymptotically the standard HSA emerges
phenomenological vertex functions are introduced, their
parameterizations bear resemblance to meson form factors. A systematic
study of all exclusive photon-nucleon reactions has been performed
\cite{kro:93}: Form factors in the space-like and time-like regions, real
and virtual Compton scattering off protons, two-photon annihilations
into proton-antiproton, photoproduction of mesons. A fair description
of all the large momentum transfer data for these reactions has been
achieved, utilizing in all cases the same baryon DAs as well as the
same values for the few parameters specifying the diquarks. Due to the
occurence of vector diquarks the model provides non-zero helicity flip
amplitudes and
consequently violations of the helicity sum rule at finite $Q^2$.
Therefore, also the process $\eta_c\to p\bar p$ can be investigated
\cite{pil:93}.
The prediction for the Pauli form factor of the proton is shown in
Fig.~4. Its fair agreement with the data as well as the reasonable
value obtained for the $\eta_c\to p\bar {p}$ branching ratio ($0.38\pm0.15$)
indicates the correct size of the spin effects generated by diquarks.
\par
\indent
As a last example of our results I want to mention the electron
asymmetry in the reaction $e p \to e p \gamma$:
\begin{equation}
  \label{asy}
A_L\,=\,\frac{\sigma(+) -\sigma(-)}{\sigma(+)+\sigma(+)}
\end{equation}
where $\pm$ indicates the helicity of the incoming electron. $A_L$
measures the imaginary part of the longitudinal
($\lambda_{\gamma^*}=0$) -- transverse ($\lambda_{\gamma^*}=\pm1$)
interference. The longitudinal amplitudes for virtual Compton
scattering $\gamma^*p\to\gamma p$ turn out to be small in the diquark
model (hence $A_L^{VC}$ is small). However, according to the model, $A_L$ is
large in the region of strong Bethe-Heitler contamination (see
Fig.~5). In that region, $A_L$ measures the relative phase
(being of perturbative origin from on-shell going internal gluons,
quarks and diquarks \cite{far:89}) between the Bethe-Heitler amplitudes
and the virtual Compton ones. The magnitude of the effect shown in
Fig.~5 is sensitive to details of the model and, therefore, should not
be taken literally. Despite of this our result may be taken as an
example of what may happen. The measurement of $A_L$, e.~g.~at CEBAF, will
elucidate
strikingly the underlying dynamics of the virtual Compton process.
\vspace*{-0.3cm}
\section{Summary}
\setcounter{equation}{0}
\vspace*{-0.2cm}
\indent
There is compelling evidence for the smallness of the perturbative
contributions to the pion's and nucleon's electromagnetic form
factors. With a few exceptions the perturbative contributions to other hard
exclusive reactions will also be too small in the
experimentally accessible range of momentum transfer.
Among these exceptions are the $\pi\gamma$ transition form factor
and the decay $\Psi\to p\bar p$. For both these reactions fair
agreement between data and perturbation theory is found. In general however,
soft, higher twist contributions seem to dominate. For the pion's and nucleon's
electromagnetic form factors the overlap contribution evaluated from
plausible wave functions, seem to have the right magnitude to account
for the data.
\par\indent
Spin effects while experimentally
large in many cases, are difficult to explain in a theoretically
satisfactory way. There are many attempts to be found in the
literature but, so far, only the diquark model provides quantitative
predictions which are in fair agreement with data. Despite of this
\nopagebreak[4]
unsettled and unsatisfactory situation it is important to persist in
the attempt to understand the dynamical foundation for exclusive
processes in QCD.
%

\newpage
\begin{figure}[t]
\vspace*{2.0cm}
\[
    \psfig{figure=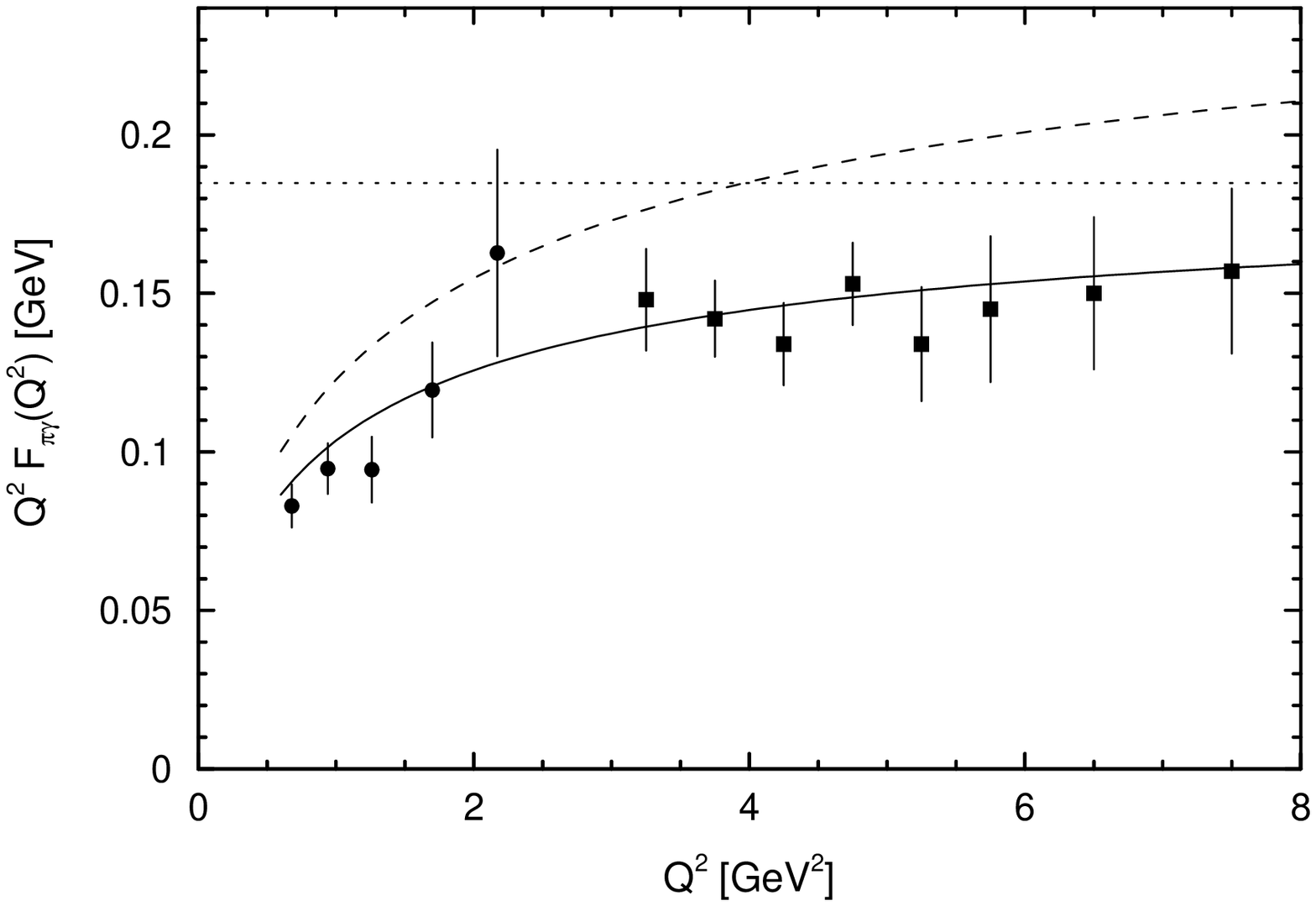,%
     bbllx=2.5cm,bblly=18cm,bburx=18cm,bbury=25cm,%
        width=8cm}
\]
\vspace*{3.2cm}

\caption[dummy1]{The $\pi\gamma$ transition form factor vs.~$Q^2$.
The solid (dashed) line represents the prediction obtained with the
modified HSA using the asymptotic (CZ) wave function \cite{jak:94}. The dotted
line
represents the results of the standard HSA (for the asymptotic wave function).
Data are taken from \cite{Beh:91,sav:95}.}
\label{fig:pigaff}
\end{figure}

\begin{figure}[b]
\[
        \psfig{figure=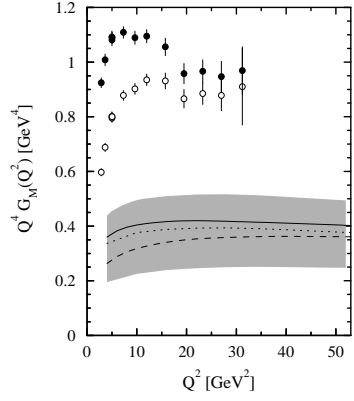,%
        bbllx=-2.5cm,bblly=18cm,bburx=18cm,bbury=25cm,%
        width=9cm}
\]
\vspace*{1.0cm}
\caption[dummy2]{The proton's magnetic form factor vs.~$Q^2$. Data are
  taken from \cite{sil:93}\\
  ( filled(open) circles: $G_M$
  ($F_1$)). The strip of theoretical results is obtained from the DAs given in
  \cite{ber:93}. The wave functions are
  normalized to unity. The plot is taken from \cite{bol:95a}.}
\label{fig:prff}
\end{figure}
\vspace*{2.0cm}

\begin{figure}[t]
\[
        \psfig{figure=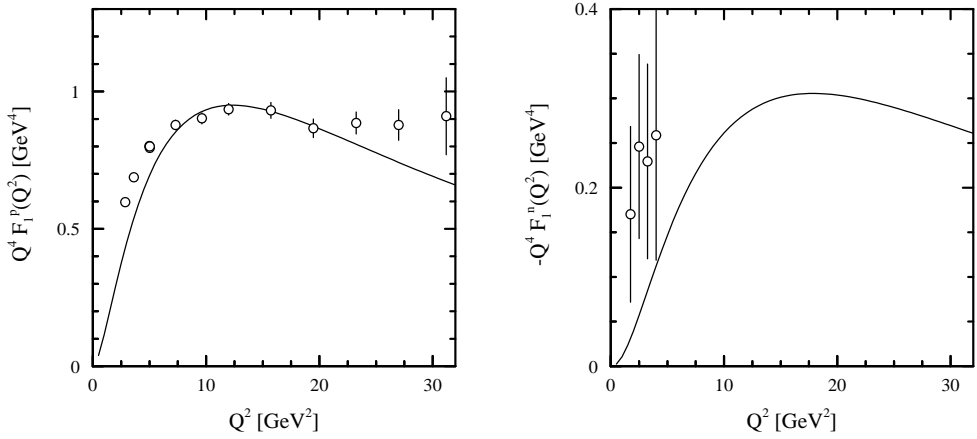,%
        bbllx=3.5cm,bblly=18cm,bburx=18cm,bbury=28cm,%
        width=10cm}
\]

\vspace*{0.1cm}
\caption[dummy3]{The overlap contributions to the proton's and
neutron's magnetic form factors \cite {bol:95b}. Data are taken
  from \cite{sil:93,lun:93}.}
   \label{fig:pff}
\end{figure}
\begin{figure}[b]
\vspace*{0.1cm}
\[
    \psfig{figure=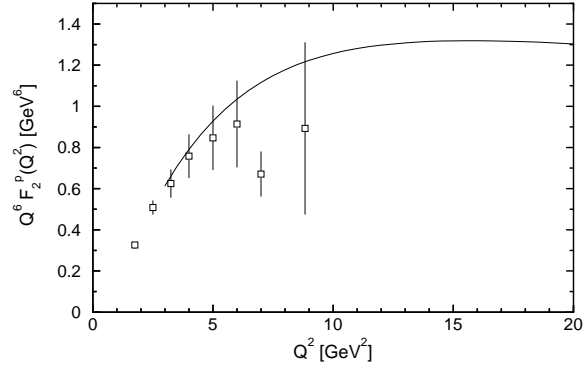,%
        bbllx=0.5cm,bblly=18cm,bburx=18cm,bbury=25cm,%
        width=8cm}
\]
\caption[dummy4]{The Pauli form factor of the proton scaled by
  $Q^6$. Data are taken from \cite{bos:92}. The solid line represents the
  result obtained from the diquark model \cite{jkss:93}.}
 \label{fig:pauli}
\end{figure}

\begin{figure}[t]
\vspace*{0.1cm}
\[
    \psfig{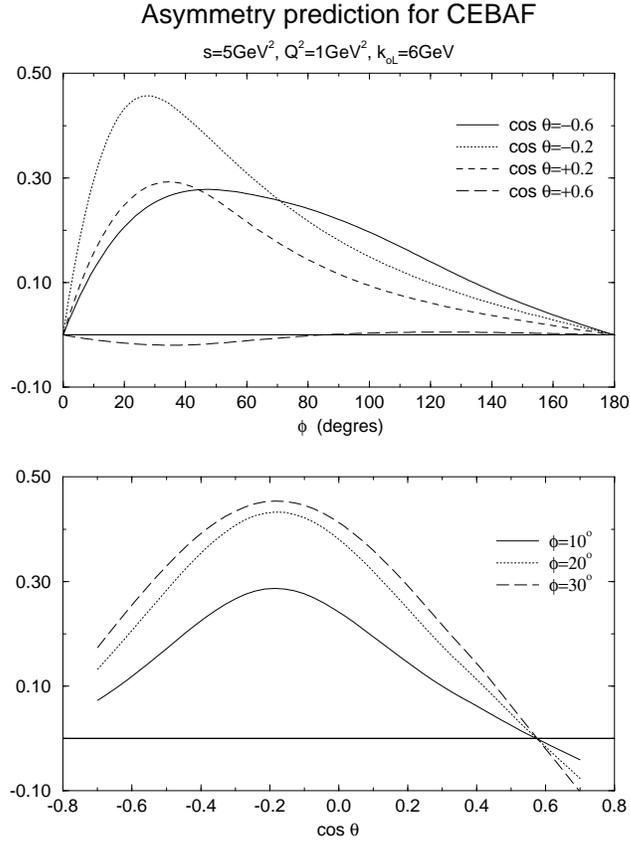}
\]
\vspace*{2.5cm}
\caption[dummy5]{The electron asymmetry in $ep\to ep\gamma$ as
  predicted by the diquark model \cite{gui:95}. $\phi$ denotes the
  angle between the hadronic and the leptonic scattering planes;
  $\theta$ is the scattering angle of the photon in the $p\gamma^*$ cms.}
   \label{fig:asym}
\end{figure}
\end{document}